\documentclass[11pt]{article}
\usepackage[margin=1.0in]{geometry}
\usepackage{graphicx}
\usepackage{epsfig}
\usepackage{amssymb,amsfonts}
\usepackage{epstopdf}
\usepackage{hyperref,color}
\usepackage{natbib}
\usepackage{setspace}

\begin{document}

\title{Time in quantum gravity}
\author{Nick Huggett, Tiziana Vistarini, and Christian W\"uthrich\thanks{Work on this project has been supported in part by the American Council of Learned Societies through a Collaborative Research Fellowship to N.H. and C.W..
Note on attribution: this paper was collaborative, though \S\ref{sec:problem} is primarily due to C.W., and \S\ref{sec:string}-\ref{sec:NCFT}. to N.H. and T.V..}}
\date{For Adrian Bardon and Heather Dyke (eds.), {\em The Blackwell Companion to the Philosophy of Time}.}
\maketitle

\begin{abstract}\noindent
Quantum gravity---the marriage of quantum physics with general relativity---is bound to contain deep and important lessons for the nature of physical time. Some of these lessons shall be canvassed here, particularly as they arise from quantum general relativity and string theory and related approaches. Of particular interest is the question of which of the intuitive aspects of time will turn out to be fundamental, and which `emergent' in some sense.

\vspace{3mm}\noindent
\end{abstract}

\noindent
As other contributions to this volume testify, physical time plays a rather different role in general relativity than in quantum mechanics and the particle physics based on it. On the face of it, quantum (and indeed Newtonian) mechanics treat time in much the same way, as a parameter presumed by, and hence independent of, dynamics. The core of the revolutions brought about by general relativity, on the other hand, was to radically reconceptualize this picture by understanding time as inextricable part of spacetime, which interacts with---and thus gets acted upon by---the physical systems it `contains'. Since a theory of quantum gravity will encompass both relativistic and quantum phenomena it must address this disparity, likely with some fundamentally new conception of time: thus candidates for such a theory promise profound philosophical lessons. Admittedly, today's candidates are provisional at best but, if history is a guide, finding the new theory will involve investigating new concepts of time arising in such steps towards a new theory, and here the philosopher can contribute. Our aims in this paper are to provide an accessible entry into this material and to assess what it suggests for the nature of time.

We will start, in \S\ref{sec:mapping}, with a brief survey of the main approaches to quantum gravity, then proceed by considering the lessons that can be drawn from two distinct strategies for discovering a theory of quantum gravity. First, in \S\ref{sec:problem} we explicate the fate of time in approaches to quantum gravity that start with general relativity and attempt to quantize it. This explication will show how successive theories down the relativistic path have shed time of many intuitive features; at the endpoint of this path we find the `problem of time'---according to an important class of theories, there is no time left fundamentally at all. 

Second, we will investigate approaches to quantum gravity that start from quantum particle physics; we offer an assessment of time as understood in string theory in \S\ref{sec:string} and in `non-commutative' field theory in \S\ref{sec:NCFT}. The former has received a great deal of attention from physicists and the public, though not so much from philosophers; the latter approach is less well-known outside of physics, though it can arise as a low-energy limit in string theory. Our focus will be on the role of causal ordering in the concept of time: in string theory there may be nothing but causal order; while non-commutative field theory raises the possibility of backwards causation.

\section{Mapping quantum qravity}\label{sec:mapping}

Before we set out to analyze the nature of time in quantum gravity, a few remarks on what quantum gravity is and why one would want to pursue a theory of it are in order. The most compelling reasons are phenomena in which both gravitational as well as quantum effects should play an ineliminable role: phenomena, such as the very early universe and the dynamics of black holes, in which high energy densities combine with strong gravitational fields so that neither can be neglected. Any adequate theory of these phenomena must coherently model the interaction of quantum matter with strong gravitational fields. While this motivation is potentially compatible with a theory in which quantum matter fields interact with a purely classical gravitational field \citep{hugcal01,wut05}, this essay will investigate more radical approaches, which aim to treat gravity itself as quantum mechanical. There are three broad categories of such theories:

\begin{itemize}
\item The first road to quantum gravity starts out from our best classical theory of gravity, general relativity (GR), and undertakes to subject it to a so-called `quantization' procedure, i.e., a mathematically rigorous transposition of the classical theory into a quantum theory.\footnote{It should be noted that the classical theory is prior to the quantum theory only in the context of discovery, not in the context of justification.} The promise of such quantization procedures to obtain a quantum theory of gravity derives from their success elsewhere in physics, e.g., in turning classical electrodynamics into quantum electrodynamics. This family consists essentially of two genera: the somewhat inactive covariant approach and the more vigorous canonical camp. In order to apply a canonical quantization scheme, the classical theory must first be cast in a Hamiltonian formalism, a procedure that involves the occurrence of `constraints'. These lead quite directly to some perplexing---and profound---philosophical questions regarding time, change, and the dynamics of a physical system, as will be discussed in \S\ref{sec:problem}. The most important---but not sole---representative of this family (and the canonical genus) is loop quantum gravity (LQG).

\item The second group takes quantum theories of matter as its vantage point and seeks to generalize or extend them to incorporate an account of strong gravitational fields and their interactions with matter fields. Attempts to assimilate gravity to the other forces of nature within the framework of the `standard model' were frustrated by the apparent `non-renormalizability' of the theory. As a result, physicists attempted other approaches, which aim to derive all forces, including gravity, as a low-energy limit of some kind. String theory is by far the most important instance, but there are others, such as non-commutative geometry; these approaches are investigated in \S\ref{sec:string} and \S\ref{sec:NCFT}, respectively.

\item The third family consists of truly iconoclastic approaches that venture to replace large swathes of `old' physics by radically novel principles and ideas in order to articulate a theory of quantum gravity {\em ab initio}. Members of this family tend to offer programmatic schemes rather than full-fledged theories. They gain in attraction as more conventional approaches fail to produce a complete and coherent quantum theory of gravity. Even if none of them ultimately work out, they provide ample opportunities to consider the implications of, and interactions among, fundamental principles which may (or may not) feature in a final quantum theory of gravity. Somewhat arbitrary examples of this category include causal set theory and causal dynamical triangulation theory.
\end{itemize}

All of these approaches have their share of problems and challenges; especially, each has at best a remote and tenuous connection to experiment, and so there are only weak empirical constraints on theory construction and choice. All the same, despite its speculative nature, we nevertheless believe that studying quantum gravity is a rewarding philosophical activity, as we will now begin to demonstrate.

\section{Quantizing GR: The problem of time and its origin}\label{sec:problem}

In the fifth century BCE, the ancient Greek philosopher Parmenides maintained that the fundamental reality is ever unchanging. What exists is, at core, completely `frozen' in time. All the change that we perceive, including the ephemeral nature of our experiences itself, only occurs at the superficial level of appearances. Furthermore, these appearance are, according to Parmenides, illusory and deceive us about the nature of fundamental reality. Although most philosophers since antiquity have found this thesis ludicrous, surprising results in theoretical physics now appear to vindicate Parmenides after all: the canonical approach to quantum gravity, which includes LQG and is described in \S\ref{sec:mapping} as one genus of the first family of approaches, apparently implies that there cannot be any change over time, fundamentally speaking. In a stark contrast to this evaporation of change, our phenomenal experience seems to mandate an account of physical reality which involves, or at least permits, change in what there is and in how things are. This is the so-called `problem of time' in quantum gravity. While that label is often awarded to a number of related, but subtly differing issues, one can readily discern at least two basic aspects of the problem: the disappearance of time as a fundamental magnitude and the freezing of the dynamics. Let us address these aspects in turn.

\subsection{Time in classical spacetime physics}
The first issue concerns the objective physical existence of time. This is not meant to necessitate that time be an Aristotelian substance, but only that our best theories in fundamental physics require a time parameter with respect to which the dynamics unfolds as postulated by the theory. Quite independently of whether time is regarded as a substance in itself or only as arising from appropriate relations among material objects or physical events or as a dynamical parameter, most would agree that it {\em orders} what we take to be the basic constituents of the universe. For instance, it tells us which of two (if any) events or processes or states of affairs occurs earlier. Specifically, one would expect time to be captured by, or give rise to, a binary relation, which orders---{\em temporally}, of course---a set of, say, events. A very natural and robust notion of time could be maintained if the following two conditions could be ascertained of this relation: (1) it partitions the set of events into equivalence classes of simultaneous events, and (2) it defines a total order on the set of (equivalence classes of) events.

Denoting the set of events by $\mathcal{E}$ and the binary relation by $\leq$ (read `precedes'), we inveterately require (1), or, equivalently, that $\leq$ induce an equivalence relation naturally interpreted as simultaneity. The idea behind this requirement is that every event $e$ in $\mathcal{E}$ should belong to exactly one subset of $\mathcal{E}$ consisting of all events simultaneous to $e$. As the symbol $\leq$ already suggests, the binary relation interpreted as temporal precedence is defined, without loss of generality, to be reflexive and so also to hold of pairs of simultaneous events (and should thus be read, more correctly, as `precedes or is simultaneous to'). Thus, $\leq$ induces a binary relation $S$ such that $\forall x, y \in \mathcal{E}, Sxy$ iff $x\leq y$ and $y\leq x$. It is easy to convince oneself that $S$ is reflexive, symmetric, and transitive and hence an equivalence relation. Thus, the set of events $\mathcal{E}$ is partitioned into sets of simultaneous events.

Because we wish not only to determine which pairs of events occur simultaneously, but also which of any two non-simultaneous events precedes the other, we naturally impose more structure of $\mathcal{E}$. We do this by demanding that, somewhat more formally, the `quotient set' $\mathcal{E}/S$, i.e.\ the set of all sets of simultaneous events, be `totally ordered'. A binary relation $R$ defines a {\em total order} on a set $X$ just in case for all $x, y, z \in X$, the following four conditions obtain: (1) $Rxx$ (reflexivity), (2) $Rxy \& Ryz \rightarrow Rxz$ (transitivity), (3) $Rxy\& Ryx \rightarrow x=y$ (weak antisymmetry), and (4) $Rxy \vee Ryx$ (comparability). Bearing in mind that the relata of the total order are not events in $\mathcal{E}$, but entire equivalence classes $\mathcal{E}/S$ of simultaneous events, it is straightforward to ask $\leq$ to be a total order of $\mathcal{E}/S$.\footnote{The reader should pause to convince himself or herself that this is indeed straightforward.} In other words, the sets of simultaneous events are linearly ordered such that $\leq$ easily permits a parametrization (with parameter $t$) enshrining that linear order.

Time as it occurs in classical Newtonian physics carves up---{\em excusez l'anachronisme}---`spacetime' into sets of simultaneous events which it totally orders. In fact, time in Newtonian theories has additional structure; more specifically, it has a metric determining not just the ordering among events, but also the duration between moments. These theories are considered by some to be hospitable to the grafting on of a more meaty metaphysics of time involving an objectively privileged spatially extended present or objective temporal becoming. 

Starting with special relativity, formulated by Albert Einstein in 1905, subsequent developments in physics have taken their turns to peel away the structure of time as present in Newtonian theories. Special-relativistic theories deny that all pairs of events are related by $\leq$: spacelike related events do not exemplify it. This entails that the set of events can no longer be totally ordered; it is merely `partially' ordered. A {\em partial order} on a set is defined as a binary relation which satisfies the first three, but not the fourth, conditions of a total order, thus admitting pairs of elements which fail to exemplify the binary relation. This loss of comparability is entailed by the loss of absolute simultaneity in special relativity: there simply is no frame-independent fact of the matter as to which of two spacelike-related events precedes the other. Of course, this can be remedied by considering matters from a particular reference frame; but unless there is an objective way to privilege this frame, such a move makes the temporal relation ternary rather than binary, taking as one of its relata the particular reference frame considered.

A further layer of structure is shed once we move to general-relativistic theories. Here, as a consequence of GR's topological permissiveness, a spacetime may contain causal loops.\footnote{Cf.\ \citet{smewut11}.} This entails that the weak antisymmetry, and thus the partial ordering, is lost.\footnote{Because an event can now precede {\em and} be preceded by another without being simultaneous to it.} If the binary relation remains only reflexive and transitive, but neither weakly antisymmetric nor comparable, then it defines a {\em preorder}. Thus, pairs of events can stand in no temporal relation to one another or in a temporal relation both ways so that they both precede one another without being simultaneous. Generally, then, there exists no globally defined time in GR which would justify the sobriquet. That GR offers a rather inhospitable home to our usual concept of time emerges even more starkly when we move from its standard articulation to its so-called Hamiltonian formulation. This reformulation is necessitated by the canonical quantization used for an important class of quantum theories of gravity, as we have seen in \S\ref{sec:mapping}.

In order to appreciate the `problem of time' as it arises in Hamiltonian GR, a central foundational aspect of GR---its so-called `background independence'---must be explicated. The pre-relativistic equivalence principle demands the equivalence of inert and gravitational mass of all physical bodies. In GR, this principle is generalized to command that effects arising from gravitational fields and those produced by inertia are caused by one and the same structure, the so-called {\em inertio-gravitational field} usually taken to be described by the metric field $g$. This inertio-gravitational field gets dissected differently into gravitational and inertial components in different frames, just as was the case with the electro-magnetic field in electrodynamics. This contrasts sharply with Newtonian physics, where a variable gravitational field is placed on a fixed spatio-temporal background. This spacetime background is what gives rise to inertial effects. This background is fixed in that it affects the physics unfolding on it without being affected by that physics. The generalized equivalence principle no longer permits a fixed inertial background and thus renders the theory `background independent'.\footnote{For a more rigorous account of background independence, cf.\ \citet{bel11}. Belot argues that whether or not a theory is background independent depends on its interpretation.} Background independence gets formally implemented in GR as the demand that the theory be `generally covariant'. Following \citet[776]{ein16}, this amounts to postulating that the laws of nature be expressed by equations valid for any coordinate system. In a modern articulation, one demands that spacetime diffeomorphisms map models of the theory to models of the theory, i.e.\ that a theory's set of models is closed under spacetime diffeomorphisms.\footnote{A {\em diffeomorphism} is a smooth and invertible map from a manifold onto itself. For more details on how general covariance gets encoded in GR, cf.\ \citet[\S3.2]{wut06}.} Physically, spacetime diffeomorphisms are `gauge symmetries' in a background independent theory in the sense that they identify two mathematically distinct models as representing the same physical situation. It is this central feature of GR which weaves space and time into an inextricable unity and engenders the deep puzzles we are about to explore.

\subsection{Time in quantum general relativity}

Canonical quantization is among the most successful recipes for cooking up a quantum theory from a classical theory. In order to make a theory amenable to the procedure, it must be cast in a so-called Hamiltonian formulation. A Hamiltonian formulation interprets the physical system it describes as a spatially extended system which dynamically evolves over time. It presupposes the existence of an external, `fiducial' time. Thus, a dynamical interpretation is foisted upon GR, forcing space and time asunder.\footnote{Accordingly, \citet[9]{mau02} complains about GR being forced ``into the Procrustean bed of the Hamiltonian formalism.''} It is no surprise then that the dynamical equations of the Hamiltonian formulation alone are not equivalent to Einstein's field equations of the standard formulation. In order to (almost) achieve this equivalence, additional `constraint equations' must be imposed on the Hamiltonian system. These constraint equations are the mathematical expression of the presence of the aforementioned gauge symmetries which welded space and time together and can thus be understood as the way in which the generalized equivalence principle gets encoded in Hamiltonian GR. One of the constraint equations deserves particular attention. In the standard lore about Hamiltonian systems, the Hamiltonian function $H$ captures the total energy of the (closed) system at stake and generates the system's dynamics via the Hamilton equations. In Hamiltonian GR, it turns out that $H$ itself figures in a constraint equation, which sets it (weakly) to zero. The absence of an external time with respect to which the spatial system could evolve thus leads to the vanishing of the generator of the dynamics, dynamically `freezing' the system. Hence, at least one version of the `problem of time' arises already at the purely classical level.\footnote{Cf.\ \citet{cal10} for an elementary account of the problem of time and \citet{ric06} for a survey of the philosophical discussion and some proposals to resolve it.}

Following the canonical quantization procedure as introduced by Paul Dirac, one elevates the canonical variables of the classical theory to quantum operators acting on a Hilbert space which satisfy the canonical commutation relations. Classically, the constraint equations set some functions of the canonical variables to zero. In the quantum theory, they thus turn into wave equations with operators, which are functions of the basal operators, acting on the quantum states and annihilating them. Accordingly, the Hamiltonian constraint $\hat{H}$ acts on quantum states $|\psi\rangle$ so that
\begin{equation}\label{eq:hamcons}
\hat{H} |\psi\rangle = 0.
\end{equation}
Only those quantum states which satisfy these and all other constraint equations are considered physical. Jointly, they constitute the physical Hilbert space of all possible states of the quantum system. Since the constraint operators generate the gauge symmetries of the theory, every operator encoding a genuine physical magnitude (`observable') has to commute with them. But since the Hamiltonian $\hat{H}$ is also a constraint operator, every observable must commute with the Hamiltonian. Quantum observables which commute with the Hamiltonian are constants of motion, i.e., quantities which are conserved through motions and thus over `time'. Essentially, this is the problem of time in its purest form: all genuine physical magnitudes do not change, but are dynamically `frozen'. Any change in their mathematical description must thus be gauge, i.e.\ pure redundancies of representation. This is Parmenides's revenge: change is only apparent, due to the vagaries of representation; fundamentally, there is no change.\footnote{At least no `B-series' change, i.e., there is no change in the properties of a physical system, but only `D-series' change, i.e., there exists a temporally ordered series of coincidence events, as \citet{ear02a} reminds us.}

Of course, that there is no change does not imply that there is no time. So the problem which was identified in the previous paragraph would more appropriately be called the `problem of change'. But to conclude that none of this challenges the status of time would be rash, for several reasons. First, we must remind ourselves that following GR, there is no such thing as an external fiducial time with respect to which we might ascertain change. Time is part of the physical system itself that we are trying to describe quantum mechanically. If anything, time will emerge as a property intrinsic to the physical system at stake, viewed, perhaps, at scales sufficiently close to those at which GR offers a successful description.\footnote{For a first stab on the emergence of time in canonical quantum theories of gravity, cf.\ \cite{butish01}.} Second, the dynamical equation (\ref{eq:hamcons}), also called the `Wheeler-DeWitt equation', formally resembles the Schr\"odinger equation of ordinary quantum mechanics (QM),
\begin{equation}\label{eq:schrod}
\hat{H}|\psi\rangle = i\hbar \frac{\partial}{\partial t} |\psi\rangle,
\end{equation}
where $\hbar$ is the reduced Planck constant, with the crucial difference that the right-hand side of (\ref{eq:hamcons}) contains nothing but a fat `zero' instead of the temporal derivative we find in (\ref{eq:schrod}). This can be read as an indication both that the time parameter $t$ is absent in the `dynamical' Wheeler-DeWitt equation (\ref{eq:hamcons}), and that temporal change does not exist. Both the absence of time as well as of change in (\ref{eq:hamcons}) have also been identified as the problem of time in quantum GR.

\subsection{Reactions to the problem of time}

Clearly, the neo-Parmenidean conclusion found in the previous paragraph flies in the face of our experience as of a fleeting temporality. As with any paradoxical result, one can choose either to accept it and to dissolve the appearance of paradox, or else to reject it and to explicate what went wrong either in at least one of the premises or in how we got from the premises to the conclusion. Transposed to the present case, the former endorses the timelessness (or at least changelessness) of fundamental physics\footnote{Or at least that the Hamiltonian constraint in canonical GR generates gauge.} and must hence explicate the emergence of time and change from a fundamentally frozen world; the latter consider the argument in the preceding paragraph a reductio and must consequently identify what they reject in the above argument and give an account of how to implement general covariance \citep[S217]{ear02b}. Examples in the physics literature of the former kind include \citet{rov11} and, somewhat more radically, \cite{bar01,bar09}; \cite{kuc93} is one of the latter type.

John \citet{ear02a} urges that both paths pose formidable challenges. He shows how a number of ways to reject the conclusion that there is no property change at the fundamental level by identifying some illicit move in the argument above fail. Alternatively, the conclusion can be resisted by conjoining additional structure to the theory if this additional structure contains observables which do change over time. Earman (ibid., \S8) concludes that attempts along these lines may well succeed---none has just yet---, but only at the price of introducing radically novel interpretations of `old' physics or new substantive physics altogether.

For those who accept the conclusion, learning to live without property change, for Earman, involves two tasks. First, they must show how we can do physics in a world fundamentally deprived of the resources for change. This becomes particularly pressing in the case of QM, where numerous technical and interpretational issues arise for systems---such as our universe---for which the dynamical Schr\"odinger equation (\ref{eq:schrod}) degenerates into the Wheeler-DeWitt equation (\ref{eq:hamcons}), and in statistical physics, where one faces the difficulty of explicating how thermalization can occur in the absence of time and change. Carlo \cite{rov91, rov11}'s `evolving constants of motion' programme, which aims to show how we can do physics in the absence of time and change, offers the first steps of a discharge of this task, but many questions remain open.  Second, an explanation of our phenomenology of temporality must be offered. Specifically, such an account must explain how the humans we are come to have perceptions as of change while living in a genuinely changeless world. Seemingly underestimating just how radical the consequences of the fundamental absence of change are, Earman insists, quite correctly, that ``[i]nsofar as a physical explanation must be couched in terms of genuine observables, the sought after explanation cannot be purely physical.'' \citep[21f]{ear02a} This entails neither dualism nor idealism concerning property change. But it surely poses a serious challenge to physicalism.

Earman concludes that there is a real problem of time or change, viz.\ that a forceful line of argument arising from well-established theories of contemporary physics seems to suggest that there is indeed no property change, and perhaps no time altogether. Earman's compelling case for such a radical conclusion has incited surprisingly little protest among philosophers. Two philosophers who did react promptly and forcefully are Tim \citet{mau02} and Richard \citet{hea02,hea04}. After having shown (or reminded us, depending on your perspective) that there can be B-series change in the standard formulation of GR, Healey tries to overcome Earman's radical conclusion and obtain an explanation of how change can emerge from a fundamental reality which Healey agrees does not change. Healey maintains that the same conclusion translates into the Hamiltonian formulation, as long as we accept that the physical content of the theory is not exhausted by the Dirac observables and that change must properly be characterized with respect to a frame of reference (a stance already paying dividends for special relativity), and hence cannot be captured by fundamental properties which are ex constructione frame-independent. It is curious that \citet{hea04} does not mention Carlo Rovelli's notion of `partial observables' \citeyearpar{rov02,rov11}, a connection also noticed by Dean \citet[337]{ric08}. Rovelli's approach attempts to construct gauge-invariant observables from correlations among `partial', but gauge-dependent ones. How exactly Healey's `framing' scheme relates to Rovelli's approach, how they evade the strictures of Earman's chilling conclusion, and whether approaches of this kind are best interpreted in a structuralist vein, as \citet[\S 6.6]{ric06} suggests, must remain the topic for another day. Let us conclude this section with a discussion of Maudlin's opposition instead.

\citet{mau02} discerns two separate arguments in Earman to the effect that there is no property change in Hamiltonian GR. The first one turns on the perceived need to ``sop up'' indeterminism (to use Earman's expression), while the second arises from an analysis of `observables'. Concerning the latter, as Earman emphasizes, the problem of time brings to the fore the importance of the issue of what the genuine physical magnitudes as given by GR are. Maudlin offers some perceptive criticisms of one such approach to characterizing these so-called `observables' due to Peter \citet{ber61}, complaining that they lack meaningful contact to the quantities that we do in fact observe and measure when we experimentally test GR. The point is well taken---that is exactly the challenge regarding observables, i.e., to establish the connection from the theoretically privileged observables to what is indeed observable---; however, the more standard Dirac observables (which I have introduced above) are not discussed, even though analogous issues arise.

As to the former, the argument gets traction by an antecedent understanding of the deterministic nature of the dynamics of the theory at stake. Since the gauge freedom identified above can be used to conjure up a form of indeterminism in the Hamiltonian formulation of GR, using the fact that arbitrary functions of time (and other independent variables) can be amended to any solution of the dynamical equations. This indeterminism is just gauge and hence not physically salient, or, in Maudlin's words, ``faux'' or ``completely phony''. He proposes three options of how to deal with this indeterminism. First, one can just ignore it---after all, the indeterminism is just an artefact of the representation. Pace Maudlin, this option, however, does not work if we believe that Hamiltonian GR offers the vantage point to an important approach to quantum gravity, as we should in the light of the discussion in \S\ref{sec:mapping}. Given that the Hamiltonian and the standard formulations of GR are not equivalent, and given that the Hamiltonian formulation may be vindicated by its services to quantum gravity, this resolution does not address the challenge of offering an account of the physical meaning of the theory. We should, as with any promising approach, follow it and let the conceptual and foundational chips fall where they may.

Second, one can fix a gauge and thus remove any appearance of indeterminism. No doubt this will work for many physically relevant cases. However, the problem with this approach is that if we fix a gauge, then all quantifiable properties of the state of the system will depend on the gauge chosen, and hence on the selected representation rather than on genuinely physical content. Unless the chosen gauge is somehow physically privileged, there is no reason we should take the values of physical magnitudes and how they evolve seriously. But then we would not have a genuine case of gauge symmetry. So to the extent to which we take the symmetry to be truly gauge, fixing it and reading off deep metaphysical insights is not an option.

Remains the third option, `quotienting out' the indeterminism. This is a technically involved procedure with the goal of cutting out all mathematical `surplus structure' by reformulating the theory such that all previously mathematically distinct but physically identical states get represented by one, and only one, point in the now reduced phase space. Thus, the theory is freed from gauge. This is the only principled and acceptable way of dealing with indeterminism. Maudlin laments that this will lead to the ``rather silly''---indeed ``crazy'' and metaphysically monstrous---conclusion that there cannot be change in the state of the physical system. Such lamentation, however, does not amount to a counterargument---it is a restatement of the problem! Earman's modus ponens is Maudlin's modus tollens, it seems. The force of the problem, of course, arose from taking the Hamiltonian formulation of GR seriously. To the extent to which we do take it seriously, one cannot so easily escape the grip of Earman's argument.\footnote{Maudlin is also worried that in the case of Hamiltonian GR, we may be recognizing that there is indeterminism without a prior notion of the gauge symmetries of the theory, and that the latter are only postulated after the fact to {\em render} the dynamics deterministic. But this is not so: the gauge symmetries of Hamiltonian GR can be antecedently identified in a principled way, which is exactly why this approach comes with \cite{ear03}'s recommendations.}

But the explanatory debt identified by Earman (and others) remains: if borne out, how can we humans amid a fundamentally changeless universe have so vivid perceptions as of a fleeting passage of time? As long as this explanatory debt is not discharged, any approach entailing a fundamentally frozen world comes with a promissory note. To fully redeem this note is a formidable task, both technically as well as philosophically. Presently, we simply refer the reader to \citet[\S3]{hea02} for a more detailed to-do list.

\section{Time in string theory}\label{sec:string}

The previous section investigated the consequences for time if one attempts to marry GR and QM by quantizing GR. As we explained earlier, that is only one strategy that is pursued. In this section we turn to the second major approach---\emph{string theory}---and ask whether it holds different lessons for the nature of time. 

String theory starts from quantum particle physics (or `quantum field theory'), which has successfully accounted for electromagnetic and nuclear forces in terms of the exchange of force carrying particles, such as photons. According to this paradigm, gravity is not a feature of a dynamical spacetime geometry, but a force in a fixed spacetime; thus it should be understood in terms of the exchange of force-bearing `gravitons'. Unfortunately such a theory is apparently `non-renormalizable', meaning in broad terms that it is incompatible with the interactions at a point assumed by particle physics.\footnote{Recent work has reignited a debate over the renomalizability of gravity, e.g., \cite{Wei:09}.} Thus in string theory, point particles are replaced by 1-dimensional `strings', whose extra dimension allows interactions to be `spread out' in spacetime, avoiding the problem of non-renormalizability (see \cite{wit96} for more). In this sense string theory is an elaboration of quantum field theory; as we shall see it also incorporates many of its standard techniques.

We will start with a basic presentation of the physics of strings (not even reaching the interactions just mentioned). Our presentation will sketch the tiny fraction of the formalism necessary for the points we wish to make about time. For a more comprehensive discussion we recommend \citealt{Gre:00} at a popular level, or \citealt{Kat:07} and \citealt{Zwi:09} for more advanced treatments (our presentation is based on the last two). The basic physical picture will enable us to explicate an interpretation in which time `lives' on the worldsheet of a string; then, because of the symmetries of the theory, while the temporal order (in the sense of special relativity) is well-defined, intervals of time have no well-defined durations, even in the sense of relativity.

So consider a classical relativistic string, an object of one spatial dimension and one temporal dimension. Let's suppose our string is free, subject to internal tension but under no influence from external forces; let's also suppose that it lives in $d$-dimensional Minkowski spacetime. What do we expect this $2$-dimensional object to look like? Like a spring tossed in the air (ignoring gravity) it can have an overall velocity and it can wobble, stretch and compress, so it should satisfy a simple wave equation.

\begin{figure}
\begin{center}
\includegraphics[width=3in]{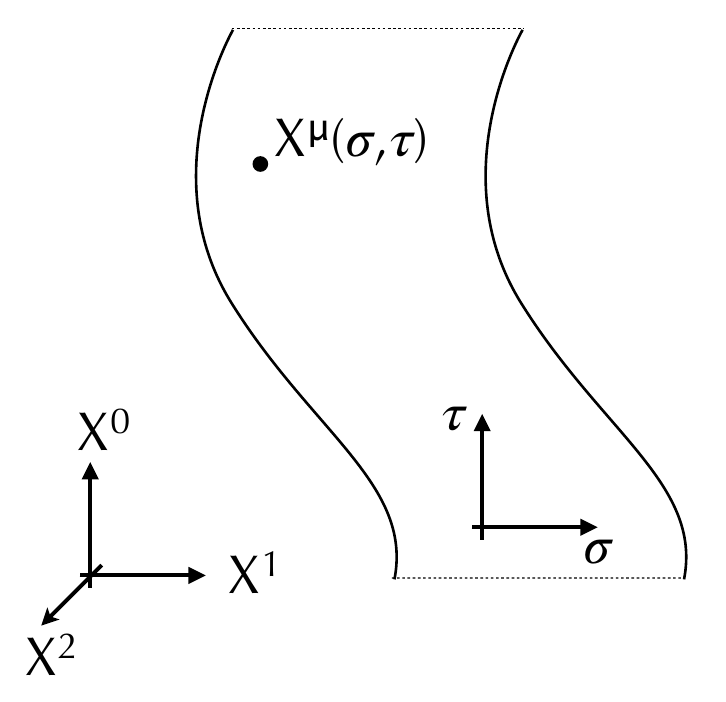}
\caption{Any point on the string has internal, worldsheet coordinates $(\sigma,\tau)$ and  external, spacetime coordinates $X^\mu$. Hence there is a function from points of the string to points of spacetime---formally, a field on the string.}
\label{fig:string}
\end{center}
\end{figure}

To write one down,  label the points of the string with `internal' space and time coordinates, $\sigma$ and $\tau$, respectively. The points of `external' Minkowski spacetime, we label with coordinates $X^{\mu}$ $(\mu= 0,1,...,d-1)$, where $X^0$ represents the time coordinate. Each point of the string occupies a point of spacetime, and we describe the configuration of the string worldsheet in spacetime by specifying the coordinates of each internal point. Formally then, a world sheet is represented by a function from the $(\sigma, \tau)$ to the $(X^{0}, X^{1},...,X^{d-1})$---in other words $X^\mu(\sigma,\tau)$ is a $d$-component vector \textit{field} on the string. From the point of view of the string, to say that the string is `wobbling' is to say that the components of this $X$-field oscillate on the worldsheet, with respect to the internal components. See figure \ref{fig:string}.

Defining $ \ddot{X}^{\mu}\equiv\partial^2 X^{\mu}/\partial \tau^2$ and $X''^{\mu}\equiv\partial^2 X^{\mu}/\partial \sigma^2$ (i.e., second derivatives of $X^\mu$ with respect to $\tau$ and $\sigma$), the simplest relativistic wave equation takes the form (with $c=1$)

\begin{equation}
\label{eq:wave}
\ddot{X}^{\mu} - X''^{\mu}=0.
\end{equation}
A generic solution  $X^{\mu}(\sigma,\tau)$  is the sum of several parts: an initial position plus velocity plus oscillatory terms, matching our intuition about the string's motion. The various modes---i.e., vibrations compatible with boundary conditions at the ends of the string---have mass and (in QM) represent particles, including the graviton.

String theory draws heavily on the techniques of quantum field theory, which requires the  `\emph{action}' rather than an equation of motion. Classically, the action of a system is simply a function, $S$, associating a special quantity with each trajectory through phase space---with a kinematically possible evolution of the system. Lagrange's `least action' principle says that a trajectory is dynamically possible if it \emph{minimizes} the action---any close variant of the path has a greater value for $S$.

In classical string theory the action can be interpreted (up to a constant, $T$, with the units of tension) as the spacetime `area' of the string:

\begin{equation}
\label{eq:NG}
S = -T\int_{\sigma,\tau} \mathrm{d}A.
\end{equation}
where dA is the area\footnote{Briefly, $\mathrm{d}A=\sqrt{-\mathrm{det}[g_{\mu\nu}\partial_\alpha X^\mu\partial_\beta X^\nu]}\cdot\mathrm{d}\sigma\mathrm{d}\tau$. Note that $g_{\mu\nu}\partial_\alpha X^\mu\partial_\beta X^\nu$ is the metric induced on the string by the spacetime metric---it is to be distinguished from the internal metric introduced below. (\ref{eq:NG}) is known as the `Nabu-Goto' action after the scientists who first explored its properties in detail.} of a region $\mathrm{d}\sigma \times \mathrm{d}\tau$. Thus the least action principle gives stringy dynamics an elegant interpretation: a string will follow the path of least area between given initial and final states. It's easy to show that this rule entails our wave equation (\ref{eq:wave}).

In classical mechanics the dynamics is obtained by minimizing the action but, as Richard Feynman discovered, the action plays a different role in QM: it's essentially the logarithm of the probability amplitude for a path. So to find the probability that a system will transition from one state to another one first takes $e^{iS}$ for each path connecting them, and sums over them to yield the amplitude; finally one squares the result to obtain the probability. Let's apply this to strings: as we saw, a possible `path' for a string is a field, $X^\mu(\sigma,\tau)$, specifying an embedding of the string in spacetime. 

In QM (\ref{eq:NG}) is intractable, so an equivalent, `sigma' (or Polyakov), action is used. This reformulation uses the trick of introducing an `internal' metric tensor, $\gamma(\sigma,\tau)$ on the string. $\gamma$ assigns lengths to curves on the string, but it is important to emphasize that these lengths are \emph{not} the lengths assigned by the big spacetime metric---indeed, as we shall see, because $\gamma$ is a formal device it has a very different physical significance from familiar metrics. The upshot is that the amplitude for a path $X^\mu(\sigma,\tau)$ with internal metric $\gamma(\sigma,\tau)$ is $e^{iS[X,\gamma]}$ and the transition amplitude is given by summing (actually integrating) this quantity over all possible worldsheets connecting the end states.

The  sigma action has several symmetries. First, because the string lives in Minkowski spacetime it is invariant with respect to Poincar\'{e} transformations of the spacetime coordinates, $X^\mu$. Second, it is invariant under reparametrizations of the string coordinates. This symmetry reflects the fact that the string is a true 2-dimensional object, not an `enduring' 1-dimensional one---a curve of constant $\sigma$ no more tracks a stringy point over time than a curve of constant $X^{i}$ represents the history of a spatial point in spacetime.

Such diffeomorphism symmetry is familiar in spacetime theories, but a third symmetry of the action is much more striking, viz.\ rescaling of the metric, or `Weyl symmetry':

\begin{equation}
\label{eq:weyl}
\gamma'(\tau,\sigma)= e^{\omega(\tau,\sigma)}  \gamma(\tau,\sigma),
\end{equation}
for any smooth function $\omega(\tau,\sigma)$. This symmetry reflects the purely formal character of $\gamma$---the length assigned to any curve has no physical significance, since it can be rescaled to anything one chooses.\footnote{On the other hand, in conjunction with diffeomorphism invariance, it means that the string is a `conformal field theory', which has very important consequences for the formal articulation of the theory.}

Weyl symmetry will be very important to the discussion of time in the remainder of this section, but first we need to introduce a popular (amongst string theorists) way of interpreting the theory, as articulated e.g.\ by Edward Witten:

\begin{quotation}
\dots one does not really need spacetime any more; one just needs a two-dimensional field theory describing the propagation of strings. And perhaps more fatefully still, one does not have spacetime any more, except to the extent that one can extract it from a two-dimensional field theory. \cite[28]{wit96}
\end{quotation}

We saw before that the action looks like that  for a $d+1$-dimensional field theory on a 2-dimensional surface, the string. Witten's proposal is to take that appearance    at face value as the ontology of the theory, \emph{instead} of an ontology of a string in an external spacetime. Such a view is a form of relationism since it reduces the space and time of experience to the spatiotemporal properties of material points, those of the string; and perhaps it is no more radical than other forms of relationism. This is not the place for a detailed consideration of Witten's interpretation; its currency among physicists justifies our interest.\footnote{\label{ft:witt}We note Witten has not banished the spacetime metric; it  appears in the action as an inner product on the $X$-field, as a fixed structure put in by hand, not a dynamical variable. So while it is true that Witten's interpretation avoids the need for a `background spacetime', it does not remove dependence on the metric as a `background object'. Thus it does not answer the critics of string theory who charge that it ignores the most important lesson of GR, the dynamic nature of the geometry.}

The interpretation denies that  big space and time---our space and time---feature in the fundamental ontology, but somehow arises from the fundamental field on the string. The relative motions of string vibrations (including the graviton modes) are presumably part of the story, but so is the internal spatiotemporal structure of the string, represented by the internal spatial and temporal coordinates. Is time (and space) fundamental after all, but in the guise of string time, $\tau$? To support this idea note that in the action the `time' integral is carried out with respect to $\tau$ (see equation (\ref{eq:NG}), but the same is true of the sigma action), so it appears that $\tau$ plays a familiar role of time.

Such a conservative view would still leave us with interesting questions about how phenomenal time, as it appears in classical spacetime theories, emerges from stringy time. Here, however, we want to emphasize that string `space' and `time' cannot be simply thought of as like ordinary space and time, just in two dimensions. The issue is  Weyl symmetry (\ref{eq:weyl}): the action is invariant under arbitrary local rescaling of the string metric. The usual interpretation of such a symmetry is that differences between models of the theory related by the transformation are undetectable, and hence unphysical---they are merely alternative representations of the same situation. (Compare with the case of constant boosts in velocity in classical, Galilean-invariant mechanics.)

To think through the significance of this point, first consider a spatial geometry with Weyl symmetry. The length of any curve is the integral of the infinitesimal line element, $\pm\sqrt{|g_{\mu\nu}\mathrm{d}x^\mu \mathrm{d}x^\nu|}$ along the line. So by choosing $\omega$ appropriately in (\ref{eq:weyl}) any curve can be chosen to have any length whatsoever. A space (or timeline) with conformal symmetry really is `metrically amorphous' in a sense even stronger than \cite[12ff]{Gru:68}; any specific metric appearing in a model of the space is simply an artifact of the representation. 

But the string worldsheet is a \emph{spacetime}. Importantly, since $e^\omega$ is strictly positive, the sign of any line element is unchanged by a Weyl transformation: the division into spacelike, timelike and lightlike paths---the causal structure of the string---is preserved. So, more carefully, on a string the duration of any timelike curve and the length of spacelike curve can be transformed to any value by a Weyl transformation. To emphasize this fact on the string, note that Weyl symmetry means (for suitable topologies) that the string metric can always be chosen to be flat (and hence set to the Minkowski metric using diffeomorphism symmetry)---not because the string is flat with respect the big spacetime, but because the choice of a particular intrinsic metric has no physical significance beyond its causal implications.

And so $\tau$ on the string---the fundamental `time' in Witten's interpretation---is quite unlike familiar time. In relativity, the time between two events depends on the frame of reference, but it is still the case that a timelike curve has a determinate duration, known as its `proper time'---not so in a theory with Weyl symmetry. (One could insist that there was `one true metric', but that would be like saying there is `one true rest frame' in relativity.) Let's operationalize the point for emphasis: imagine you had found some periodic physical process that you thought could serve as a clock, ticking off equal units of $\tau$ each cycle, wherever and whenever the process is instantiated. By Weyl symmetry, any cycle could have any duration, and any pair of cycles could have durations in any ratio at all; in other words, no such processes mark off equal periods of time. What's going on? Weyl symmetry says that if there is a periodic process, then without there being any physical change, it can take any stringy time whatsoever. Compare that to classical mechanics: a pendulum can't have a different period unless its mass or length changes.

To summarize: formally, string theory describes a $d$-dimensional field $X^\mu$ and metric $\gamma$ defined over the points of a string, $(\mu,\nu)$. Witten suggests that this description represents the fundamental ontology of the theory, in which case it seems that the fundamental nature of time is represented by the temporal properties of the metric. But because it only serves an auxiliary role $\gamma$ is subject to Weyl symmetry: so while there are facts about causal order on the string, durations (and lengths) are not physically meaningful. Once again, we see that an crucial aspect of the intuitive conception of time vanishes in quantum gravity---given an admittedly controversial interpretative assumption that we do not take a stand on.\footnote{We further note that the background inner product discussed in footnote \ref{ft:witt} also permits the formal definition of an induced metric on the string, one not subject to Weyl symmetry, only Poincar\'e symmetry with respect to the $X^\mu$s. We have not discussed this spatiotemporal structure at length because it seems derivative on the $X^\mu$ field, which does not represent points of spacetime according to Witten.}

\section{Space-time non-commutativity}\label{sec:NCFT}

Attempts to treat gravity as a force along the lines of the highly successful treatment of electromagnetic and nuclear forces run into trouble because of point interactions. String theory banishes point particles, but another response is to banish the point structure of spacetime itself. One way to construct a theory of this kind is to stipulate that  spacetime dimensions fail to commute: in the plane, for instance, it would seem to follow that nothing can simultaneously have well-defined $x$ and $y$ co-ordinates---and so nothing can ever occupy a well-defined point. For this reason, amongst others, physicists have studied quantum field theories in `non-commutative geometries' (though whether they solve problems of renormalizability is at best controversial). Below we shall give a very brief introduction to the idea, and then describe how such theories have quite profound implications for time: namely, they allow backwards causation and the failure of determinism.

The theory we will look at, however, does not contain gravity. But it is still an important example for our discussion, in the first place because the non-commutativity of geometric quantities is a potential consequence of quantizing spacetime, so of quantum gravity.\footnote{For instance, LQG predicts the existence of a smallest area for any surface, something prima facie incompatible with relativity, since in a suitably moving frame any minimal surface should have a smaller, contracted area. But, as \cite{RovSpe:03} point out, if the area operators associated with different frames fail to commute they will not be simultaneously well-defined, and the contraction need only hold at the level of expectation values---the average of area measurements in the moving frame will be smaller than the minimum area because sometimes they register zero. Of course, this example doesn't involve non-commutation between dimensions, but it illustrates the general point.} In the second, non-commutative fields arise as low energy limits of certain string theories (indeed some argue that non-commutativity is a general feature)---raising the specter of backwards causation and indeterminism for string theory, which \emph{is} a putative theory of quantum gravity. In fact we shall see how string theory evades these consequences, suggesting that causality and determinism are, after all, among the temporal features of strings.

So what could it mean for two dimensions to fail to commute? To start with a simple example, consider the plane. Picture a Euclidean rectangle whose sides lie along two coordinate axes, so that the lengths of its sides are $x$ and $y$---its area is $x\cdot y= y\cdot x$. If the geometry is non-commutative such products fail to commute, i.e. $xy\neq yx$. In this case it looks like there is no sensible quantity like an area.  How can we understand such a thing? By abandoning ordinary images of geometry in terms of a literal space (such as the plane) and presenting it in an alternative, algebraic way. In fact, our example already starts to do so: even thinking about areas as products of co-ordinates uses Descartes' algebraic approach to Euclidean geometry. Once we have entered the realm of algebra, all kinds of possible modifications arise. Especially, an abstract algebra $\mathcal{A}$ requires an operation of `multiplication', $\star$, but this can be a quite general map from pairs of elements, $\star:\mathcal{A}\times\mathcal{A}\to\mathcal{A}$, \emph{which need not be commutative}. For instance, one could define `multiplication' to satisfy $x\star y - y\star x = \theta$ (a small number). (We won't go into further details here, but note that such a relation entails commutation relations between any polynomials in $x$ and $y$; these algebraic relations carry geometric information.) Such a thing is perfectly comprehensible from the abstract point of view of algebra, but it cannot be given a familiar Euclidean interpretation via Cartesian geometry. So, non-commutative geometry is a theory that seems to describe a fundamentally algebraic world, not spatial (in the ordinary sense)---there is $x\star y$ and $y\star x$ but no literal rectangle.

This story can be readily generalized to $D$ dimensions, in which arbitrary pairs of spatial coordinates fail to commute; a $\theta$ for each pair. It can further be generalized to the case of spacetime; either only the spatial dimensions are non-commuting, or time is non-commuting with one or more spatial dimensions. In the first case one can define field theories that live on the non-commutative space: for instance, scalar fields, which are polynomials in the coordinates (commuting or not), or indeed vector and gauge fields (\citealt{DouNek:01} and \citealt{sza:03b}, both of which we have followed here). Not surprisingly, since $\theta$ is an invariant area, such theories are not Lorentz invariant (\citealt{carhar+:a}, but see \citealt{Sny:47}).

Things are different in the case of  space{\em time} non-commutativity. $x\star t - t\star x = \theta$ implies: (i)  backwards causation, as Nathan Seiberg, Leonard Susskind, and Nicolaos Toumbas \citeyearpar{N.-Tou:a} show; and (ii) the failure of determinism (more specifically of `unitarity'), as discussed by Jaume Gomis and Thomas Mehen \citeyearpar{J.G:00}. Let's focus on the first case, and the scattering of massless particles considered by Seiberg and collaborators.

Consider a commutative theory, and imagine two particles scattering off each other at a point. Suppose their interaction is described by a term $g\phi\cdot\phi\cdot\phi\cdot\phi$ in the action; $g$ is a coupling constant and $\phi$ is the field. Then at the first (`tree') level of perturbation theory the scattering amplitude is $-ig$. In the non-commutative version of the theory multiplication is replaced with the $\star$-product, so the interaction becomes $g \phi \star \phi \star \phi \star \phi$. The upshot is that there are terms in the interaction which depend on the energies $p^{2}$ of the outgoing particles and on $\theta$, and the non-commutative amplitude becomes $-ig\cdot f(p^{2} \theta)$. The periodic function $f(p^{2} \theta)$ entails that the outgoing wavepacket has three parts, of which two are relevant to us: the first appears time-delayed---`\emph{retarded}'---with respect to the collision; the second is instead an `\textit{advanced}' wavepacket, appearing \emph{before} the collision. The time gaps involved depend on $\theta$ of course, but also on the energy of the particles, and so the backwards causation is not necessarily  hidden below the scale we expect classical spacetime notions to break down. Now, the collision of the particles is the cause of the scattering, since the particles only interact at that time; before and after the collision the system is free. Thus, the advanced wavepacket describes a physical process in which the effect precedes the cause---in principle by a macroscopic time interval.

Seiberg et al (8-9) assimilate this backwards action to the apparently advanced process of a rigid rod bouncing off a wall; the center of mass rebounds before it hits---because the end hits first! They point out that such a process is however impossible in relativity; rigidity implies instantaneous action. But this analogy obscures matters. In the first place, as we noted, we  already expect relativity to fail so there need not be a prohibition on rigidity . Further, the violation of causality is in an important sense stronger than that of a rigid rod in which an effect has a spacelike cause: the collision is in the \emph{absolute future} of the scattered wavepacket. (Of course if one event is in the absolute future of a second, there is a third event spacelike to both; but there is no suggestion that the scattering involves two spacelike processes.)

Now, it has long been noted that the known fundamental laws of physics are time-symmetric, and hence that any `arrow of causation' needs an explanation, plausibly in terms of non-lawful facts about the world. One response to Seiberg et al is thus to say that we are faced with a case in which the arrow breaks down; not too surprising since the process follows directly from the laws. (Or to put it another way, it is the intervals both before \emph{and} after the collision that are strange; how can causes have delayed effects?) But as Gomis and Mehen show, there is a further issue, arising at the second (`1-loop') level of perturbation theory; a failure of determinism.

They show how non-commutativity between time and space produces an action which is non-local in time---hence the state of the field at a time depends on the states at \emph{both} past and future times. This situation implies a failure of `unitarity', the kind of determinism exhibited by standard QM. Briefly, in quantum theory, time evolutions are represented by `unitary operators', mapping states at one time onto states at another, either earlier or later. QM is indeterministic insofar as states encode probabilities for measurement outcomes, but it is otherwise deterministic in the sense that the unitary operators describing evolutions are invertible: they are one-to-one maps between the present state of the system and  unique states at each time in the future or past. However, the dependence of the state at a time on states in both its future and past states compromises invertibility and hence determinism: for instance, a state can evolve into different states at a later time, depending on the future state.

So much the worse for determinism and causality, if space and time turn out not to commute? On the contrary, this result is a significant problem for the approach: first, the kind of indeterminism involved is not stochastic---the different possible outcomes do not even have probabilities assigned. Second, because QM presupposes unitarity, the result means that the quantum framework breaks down; some wholly new approach is needed. Perhaps that is right, and QM and its conception of time do have to go; or perhaps we have reached a dead-end for non-commutative spacetime. Those are question for further research in physics, not something we can answer here.

However, we want to bring all these considerations back to bear on string theory, in which space and time non-commutativity can also feature. Does string theory---a much more serious contender for a theory of quantum gravity---entail the failure of determinism and backwards causation? Apparently not\footnote{Cf.\ \citet{N.-Tou:a,N.S:}, \citet{J.G:00}.}; a brief sketch of why will allow us to pose a final question, that of the significance spacetime non-commutativity in string theory.

\cite{N.S:} impose non-commutativity between space and time by introducing a background electric field, which produces a  `deformation' of the theory in which a stringy spacetime uncertainty principle can be formulated using the string tension, T:
\begin{equation}
\triangle t \triangle x \geq \frac{1}{2\pi T}.
\end{equation}
As will be familiar from QM, such uncertainty relations are alternative expressions of non-comm\-utation; non-commutativity places bounds on how well-defined simultaneously quantities can be. In this approach the commutator is imposed `extrinsically' by the presence of a background field, so that the `external' spacetime is itself commutative; but spacetime uncertainty shows up an effective non-commutative field theory at low energies. We note however that some (e.g., Tamiaki \cite{Yon:}) argue that spacetime uncertainty is an intrinsic feature of string theory, independent of the presence of background fields. The status of non-commutativity in string theory is thus an interesting open question.

Be that as it may, the formulation allows Seiberg et al to investigate whether temporal pathologies arise in the low energy, non-commutative limit. We saw above that in non-commutative field theory the $\star$-product changes the interaction term of the action to produce indeterminism and causation into the past light cone. In the case of open string scattering however, a new feature due to the oscillation of the strings shows up. Its presence modifies the mathematical expression of the non-commutative parameter $\theta$ in a crucial way \citep[11-2]{N.-Tou:a}. Although the amplitude still acquires an acausal phase, the modification of $\theta$ covers those features of the phase that in field theory are responsible for the pathologies. In fact, in this case, the actual configuration of the physical system after collision contains only retarded terms, and no advanced terms. In other words, there is no backwards causation and determinism is saved; string theory does not, after all, have pathological temporal behavior.\footnote{We will leave discussion of the origin of this temporal asymmetry for another occasion.}

To wrap up, one of the important possibilities for time in quantum gravity is that it might fail to commute with space. We have seen something of what that might mean and how it might arise, and its relation to string theory. We have also seen some of the possible implications for the nature of time that would follow, including advanced causation, and the failure of determinism, even in the weakened sense found in QM. Clearly much more remains to be said, but we hope this section serves as a signpost to further work.

\section{Brief conclusion}\label{sec:conc}

We hope to have enticed philosophers to pay attention to contemporary developments in quantum gravity, with its promise of a bountiful philosophical harvest for the philosophy of time, but also adjacent issues concerning change, causality, and determinism. Even though the field continues to be wide open, many approaches seem to suggest that physical space, or physical time, or both, will not be part of the fundamental furniture of the world. If this is borne out, then the physical time we introduced to account for our ordinary experience will play no fundamental role in the world, and hence shouldn't in our metaphysics either. But in that case, the philosopher need not panic; instead, she should welcome the challenging work that is hereby entrusted to her: to participate, alongside natural scientists, to understand how humans can come to have the fleeting experiences we do, in a world fundamentally deprived of the resources to ground time or change or both. To help articulate an account which addresses this challenge strikes us as much more exciting than many of the more traditional metaphysical task to be found in the philosophy of time.

\bibliographystyle{plainnat}
\bibliography{Blackwellbib}

\end{document}